\RequirePackage{fix-cm}
\documentclass[10pt,3p,times]{elsarticle} 

\usepackage[ruled]{algorithm}
\usepackage{algpseudocode}
\usepackage{xcolor}
\usepackage{graphicx}
\usepackage{subfig}
\usepackage[font=footnotesize]{caption}
\usepackage{url}
\usepackage[figuresright]{rotating}

\usepackage{amsmath}

\journal{}
\usepackage{array}
\newcolumntype{P}[1]{>{\centering\arraybackslash}p{#1}}

\makeatletter
\def\ps@pprintTitle{%
 \let\@oddhead\@empty
 \let\@evenhead\@empty
 \def\@oddfoot{}%
 \let\@evenfoot\@oddfoot}
\makeatother
\begin{document}
\begin{frontmatter}
\title{On the degree of synchronization between air transport connectivity and COVID-19 cases at worldwide level}

\author[BUAA,BIGDATA1]{Xiaoqian~Sun}
\ead{sunxq@buaa.edu.cn}
\author[BUAA,BIGDATA1]{Sebastian~Wandelt\corref{cor1}}
\ead{wandelt@buaa.edu.cn}
\author[UBC]{Anming Zhang}
\ead{anming.zhang@sauder.ubc.ca}

\cortext[cor1]{Corresponding author: Tel.: + 86 10 8233 8036}
\address[BUAA]{National Key Laboratory of CNS/ATM, School of Electronic and Information Engineering, Beihang University, 100191 Beijing, China}
\address[BIGDATA1]{National Engineering Laboratory of Multi-Modal Transportation Big Data, 100191 Beijing, China}
\address[UBC]{Sauder School of Business, University of British Columbia, Vancouver, BC, Canada}

\begin{abstract}
The current outbreak of COVID-19 is an unprecedented event in air transportation. While it was known that air transportation plays a key role in the spread of a pandemic, it is probably the first time that global aviation contributed to the planet-wide spread of a pandemic, with over 500,000 reported casualties related to the infection. In this study, we perform an analysis of the role air transportation played and how the air transportation system was changed throughout the ongoing pandemic. The major decision makers were countries (governments) and airlines,  who were involved mostly in the decision and implementation of travel bans. Here, we focus on the role of countries, by analyzing the degree of synchronization between the number of cases reported in specific countries and how/when these countries reacted concerning the air transportation operations. Our study also provides a comprehensive empirical analysis on the impact of the COVID-19 pandemic on aviation at a country-level. It is hoped that the study can lead to novel insights into the prevention and control of future waves or pandemics of other types.

\end{abstract}
\end{frontmatter}

\section{Introduction}
\label{sec:intro}

The 20th century was undeniably the century of aviation in transportation, from the design of the first aircraft~\citep{ACenturyInTheSky}, the engineering of increasing effective engine technology (notably the invention and refinement of jet engines), to the exploitation of efficient hub-and-spoke networks~\citep{gillen2005regulation}, ending with predicted annual growth rates between 2\% and 5\%~\cite{janic2000assessment,lee2009aviation}. The whole aerospace industry has not only revolutionized the way we travel around the world; but also significantly changed our view on how we see the world. Nevertheless, the enhanced long-distance mobility provided by aviation~\cite{diaconu2012evolution}, has become a two-edged sword. Except from ongoing discussions that aviation is harmful to the climate and nature, there is a second catch. Air transportation has also contributed significantly to the risk of spreading diseases. In fact, it has been known that air transportation plays a critical role in the spread of contagious diseases worldwide~\cite{Brockmann2013}; typical examples include the outbreaks of SARS 2003~\cite{Likhacheva2006}, MERS 2012~\cite{Zaki2012}, and Ebola 2014~\cite{Bogoch2015}. These previous diseases have had devastating and harmful effects on their own; yet, these impact could mostly be measured at a regional level. International organizations and communities were able to cut the transmission chain early, which ended the localized epidemics before they could turn into a full pandemic. 

\begin{figure}[t!]
\centering
\includegraphics[width=\textwidth]{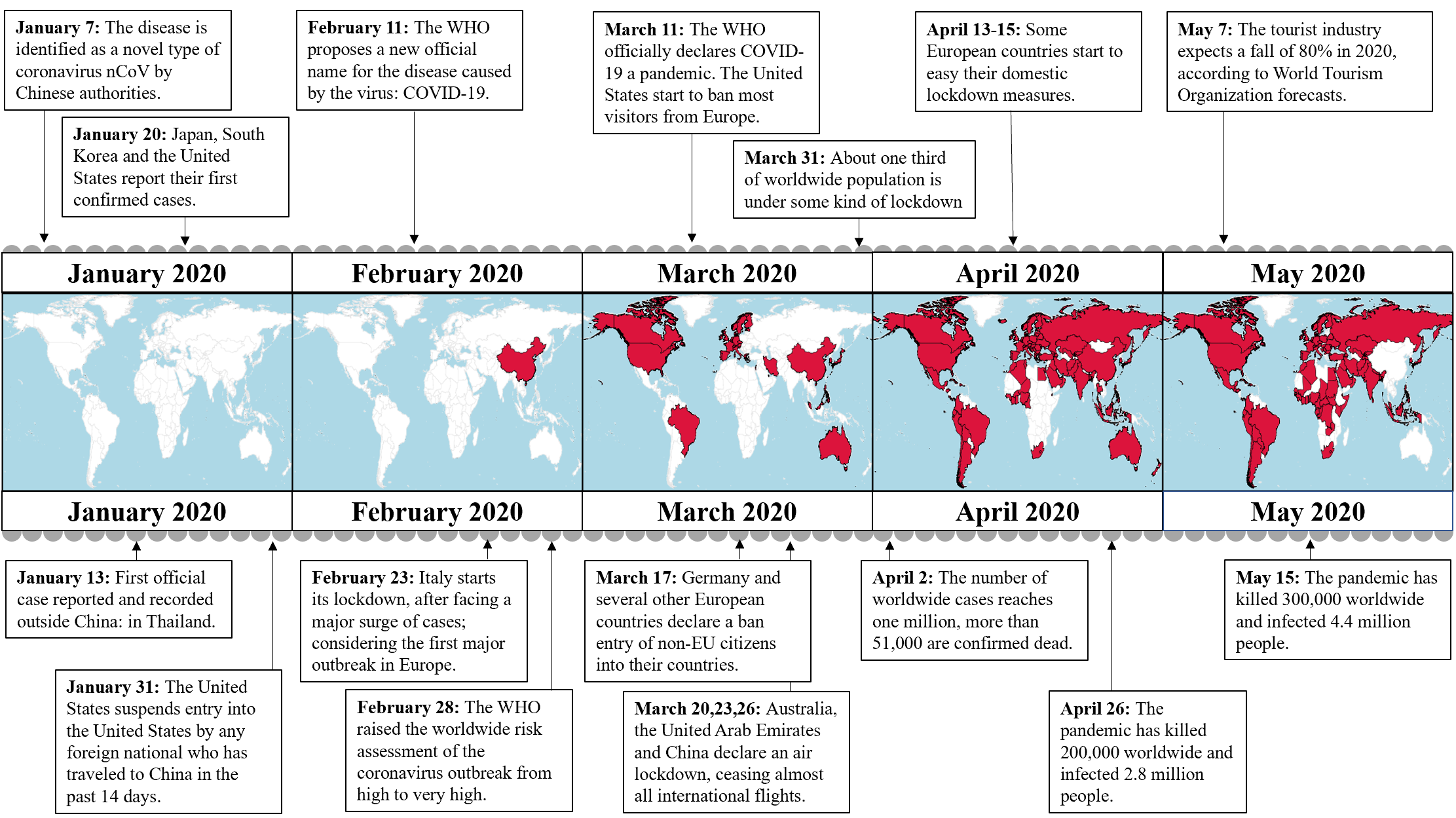}
\caption{Timeline of major events regarding the COVID-19 pandemic from January 2020 to May 2020. The five maps in the center of the figure highlight the countries which had at least 100 new infections per day on the 15th of the month.}
\label{fig:COVID19Timeline}
\end{figure}

\begin{figure}[t!]
\centering
\includegraphics[width=\textwidth]{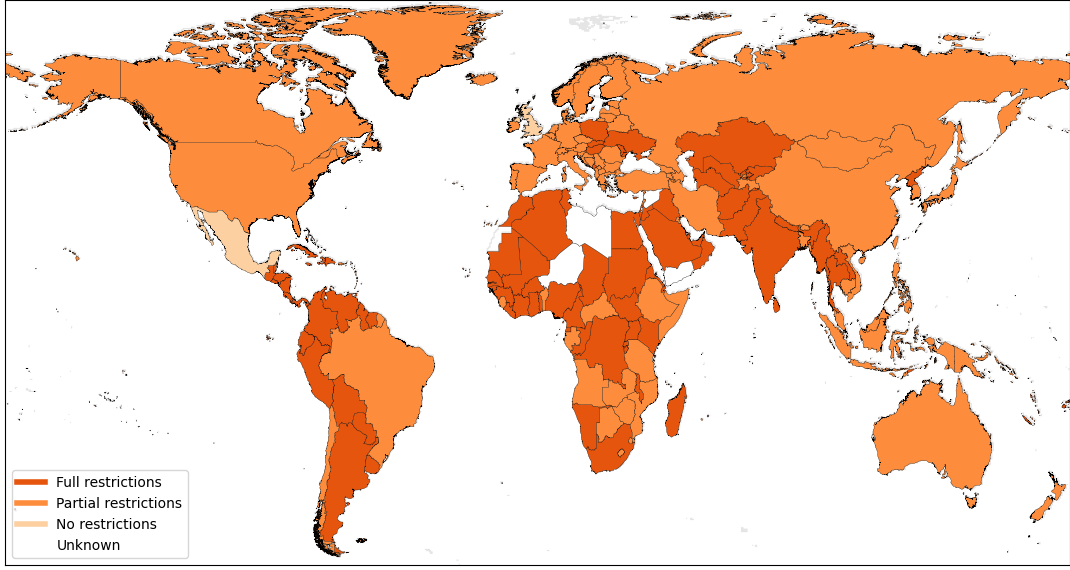}
\caption{Flight restrictions on countries as reported by IATA (May 21st, 2020). Half of all countries worldwide have full/total restrictions (in darker orange color), which means that airports are closed to all international flights/passengers. Partial restrictions (in light orange color), as implemented with the majority of remaining countries, which often accept the flights of country nationals, permanent residents (green-card holders), and other (distinct) exceptions. A few countries have no restrictions reported (in lighter orange color) or have unknown restrictions according to IATA (without color). See \protect\url{https://www.iatatravelcentre.com/international-travel-document-news/1580226297.htm} for details.}
\label{fig:Restrictions}
\end{figure}

With the emergence of COVID-19, the inconceivable, yet predicted singularity happened: Despite of heavy travel ban restrictions and quarantine policies enforced by governments, the COVID-19 outbreak, which is believed to have started around January 2020, quickly spread to almost all countries worldwide. The number of infected cases reached 10 million in late June 2020, with the disease killing more than 500,000 people worldwide. Figure~\ref{fig:COVID19Timeline} gives an overview of the major events during the pandemic outbreak from January 2020 to May 2020. Strict travel ban restrictions further amplified the substantial reduction of air passenger demand and resulted in a large number of flight cancellations. Although the total number of the COVID-19 infected cases is surging, a few countries seem to have achieved certain successes in fighting COVID-19, such as China where rebound activities have been undertaken in order to recover the economic development. Figure~\ref{fig:Restrictions} visualizes the degree of country-specific flight restrictions during COVID-19 in May 2020. We can observe that more than half of all countries worldwide still have full or partial flight restrictions and that the mobility of air passengers is heavily constrained, particularly for international air travelers. Recent studies have analyzed the role transportation systems played mainly inside China~\cite{Kraemer2020,Li2020,Chinazzi2020,cheung2020evolution}, combining air transportation 
and high-speed railway systems, which are the prevalent mid-range transportation modes in China. Moreover, some studies aimed at developing disease spreading models, which quantify the risk of disease spreading from China to the other countries~\citep{Christidis2020,Gilbert2020} and the role government policies had on reducing the number of infections and effects on economic growth~\cite{Hsiang2020}.

In this study, we address the following question: What degree of synchronization can we find between air transport connectivity and the number of confirmed COVID-19 cases? Given the known importance of air transportation on disease spreading, together with the fact that early cases of COVID-19 were known at the beginning of January, one would have expected, that the aviation system would have been (at least partially) locked down \emph{before} the disease aggressively spread around the world; travel restrictions are critical particularly at early stages of an outbreak~\cite{Kraemer2020}. The two major decision makers to be considered here are countries (governments) and airlines. Countries have the political power to protect their people by closing borders, mainly by the means of invoking travel bans on part of the population. Airlines, on the other hand, being mainly profit-driven, need to meet the demand of travellers, in order to reach sufficient load factor, to be profitable. Here, we focus on the role of countries, by trying to identify the order of events, involving the closure of country boundaries and the temporal evolution of COVID-19 cases. Given our hypothesis, one would expect that the operations of aviation would have been significantly reduced before the virus entered the country. The findings in our study, however suggest completely opposite: Most operations were ceased only long after the first cases were recorded. Our study is based on actual flight data for the period of January 2020 to mid-May 2020. Overall, this study investigates the two-way relationships between air transportation and COVID: on one hand, air transport (connectivity, or policies of cutting air transport connectivity) clearly affects the spread of COVID. On the other hand, we also investigated the impacts of COVID on aviation at worldwide level.

The remainder of this study is organized as follows. Section~\ref{sec:LiteratureReview} discusses the relevant literature on air transportation networks, with a specific focus on disease spreading. 
Section~\ref{sec:results} reports the results of our study regarding the synchronization of air transportation operations and COVID-19 cases. Finally, we summarize the major findings of the paper and discuss future research directions in Section~\ref{sec:conclusions}.

\section{Literature review}
\label{sec:LiteratureReview}

Network science tools have been shown to be extremely useful to better understand the structures and dynamics of disease spreading on complex networks, as well as studying the complexity of air transportation as a system. We discuss both of these application areas below. 

Concerning the application of network science to disease spreading, \cite{Colizza2006} presented a stochastic computational framework for the forecast of global epidemics based on the IATA database for the year 2002, focusing on the interplay between the network topological structure and the stochastic features of the infection dynamics. 
\cite{Brockmann2013} proposed a novel notion of distance, the so-called effective distance, derived from the underlying mobility network, instead of the conventional geographic distance; with the goal to investigate the hidden geometry of complex, network driven contagion phenomena. This approach was applied to the worldwide 2009 H1N1 influenza pandemic and 2003 SARS epidemic. 
\cite{Iannelli2017} proposed a random walk effective distance to predict disease arrival times on complex networks instead of the most probable path. Results showed that the generalized effective distance overcame the restriction of simple path propagation of a disease and the shortest-path measure is a particular case of the proposed random walk effective distance. \cite{Christidis2020} presented a methodology during the early stages of the COVID-19 pandemic to measure the risk of the disease spreading outside China, based on the Sabre database tracking the number of passengers from the year 2016 to 2019. The number of reported cases was compared to the estimated traffic density and an indicator of the expected rate of infection of aviation passengers was derived. While focusing on the initial phases of the pandemic, the risks associated with local infections were underestimated, and the asymptomatic carriers could not be measured given that COVID-19 has a long incubation period (up to or even exceeding 14 days). \cite{Gilbert2020} evaluated the preparedness and vulnerability of African countries against the risk of importation of the COVID-19. They found that some African countries with the highest importation risk have moderate to high capacity to respond to the COVID-19 outbreak, while those at moderate risk have variable capacity and high vulnerability. \cite{Kraemer2020} used real-time travel data from one of the biggest internet companies in China, Baidu, to investigate the impact of control measures on the COVID-19 spread in China at the start of 2020. It was shown that travel restrictions are particularly useful in the early stage of the outbreak compared with the stage that the outbreak is more widespread, and the drastic control measures implemented in China substantially mitigated the pandemic spread. \cite{Li2020} focused on the roles of undocumented cases on the overall prevalence and pandemic potential of COVID-19, using observations of reported cases in China, together with mobility data from Tencent, a networked dynamic meta-population model, and Bayesian inference. 
They estimated that approximately 86\% of all infections were undocumented before the travel restrictions implemented in January 2020, and the transmission rate of undocumented infections were around 55\% more than that of the known cases. 

\cite{Chinazzi2020} used a global metapopulation disease transmission model to study the influence of travel limitations and quarantine on the national and international spread of COVID-19. Results showed that the travel quarantine around Wuhan has only modestly delayed (3-5 days) the spread of the disease inside China, and the effects of sustained travel restrictions internationally are modest as well, unless combined with further transmission reduction interventions. 
\cite{Hsiang2020} applied reduced-form econometric methods to measure the effects of anti-contagion policies on economic growth as well as the growth rate of COVID-19 infections. They found that anti-contagion policies have significantly and substantially slowed the growth of COVID-19 infections, with the order of 62 million confirmed cases, corresponding to averting approximately 530 million total infections. \cite{Guan2020} analysed the supply-chain effects of COVID-19 control measures using a global trade modelling framework. They found that supply-chain losses to initial COVID-19 lockdowns largely depend on the number of countries imposing restrictions and the losses are more sensitive to the lockdown's duration than its strictness. 
Based on interviews with 16 senior aviation industry executives from European organizations, \cite{Suau-Sanchez2020} estimated the medium- and long-term impacts of COVID-19 on commercial aviation, focusing on three aspects: supply, demand (passenger behavior), and regulatory impacts. It was found that the future ticket prices and the future development of air freight are highly uncertain. Other studies put an emphasis on the role of multi-modal transportation impacts related to COVID-19, with a focus on China. \cite{Zhang2020} examined the factors influencing the number of imported cases, together with the speed and the spread pattern, of the COVID-19 pandemic spread in China, taking into account different modes of transportation (high-speed rails, coach, and air services). They found that the frequencies of high-speed rails and air services out of Wuhan are significantly associated with the number of the COVID-19 reported cases in the destination cities; the presence of high-speed rail stations and airports is also significantly related to the spread speed of the pandemic. 

Concerning the application of network science to the analysis of transportation networks, several studies have been published in recent years. \citep{Li-Ping2003} was among the first to find that the US airport network has small-world characteristics and its degree distributions follow two-segment power-laws. 
\citep{Xu2008} analyzed the US domestic passenger air transportation network using weighted complex network methodologies. The temporal evolution of the US airport network at city level from 1990--2010 was analyzed by \citep{Jia2014266} as well as by \citep{lin2014evolving}. \citep{Neal20135} assessed the business travel US airport network for the period 1993--2011 at node level, dyadic level, and system level. The system-wide analysis revealed that business travel among US cities is increasingly symmetric and evenly dispersed. The evolution of the European airport network between 1990 and 1998 was analyzed by~\citep{Burghouwt2001311} and the authors did not find an indication of hub-nodes for intra-European traffic. The spatial configuration of airline networks in Europe was analyzed in~\citep{burghouwt2003spatial}. \citep{Gurtner2013} analyzed the community structure in European air transportation.  Several works analyzed the airport network in Italy~\citep{Guida2007527,quartieri2008complex,quartieri2008topological}, Portugal~\citep{Jimenez2012383}, France~\citep{Thompson2002273}, as well as at the global level~\citep{cheung2020evolution}.
\citep{paleari2010} compared the structure and performance of the airport networks in the US, Europe, and China in order to find out which network is most beneficial for the passengers. The dataset was based on scheduled flights operating for October, 24th, 2007 and the data was provided by Innovata. The results showed that the Chinese airport network provides the quickest travels for passengers; the US airport network is the most coordinated; while the European airport network provides the most homogeneous level of service. A dynamic fluctuation model is proposed by ~\citep{Zhang2014590} and evaluated on the airport networks for China, Brazil, and Europe. The European air route network was analyzed by~\citep{SunTB2014}. \citep{Zanin2015} discussed the multi-layer representation of functional networks, with the European airport network as a test case.

\section{Results}
\label{sec:results}

\begin{figure}[b!]
\centering
\includegraphics[width=\textwidth]{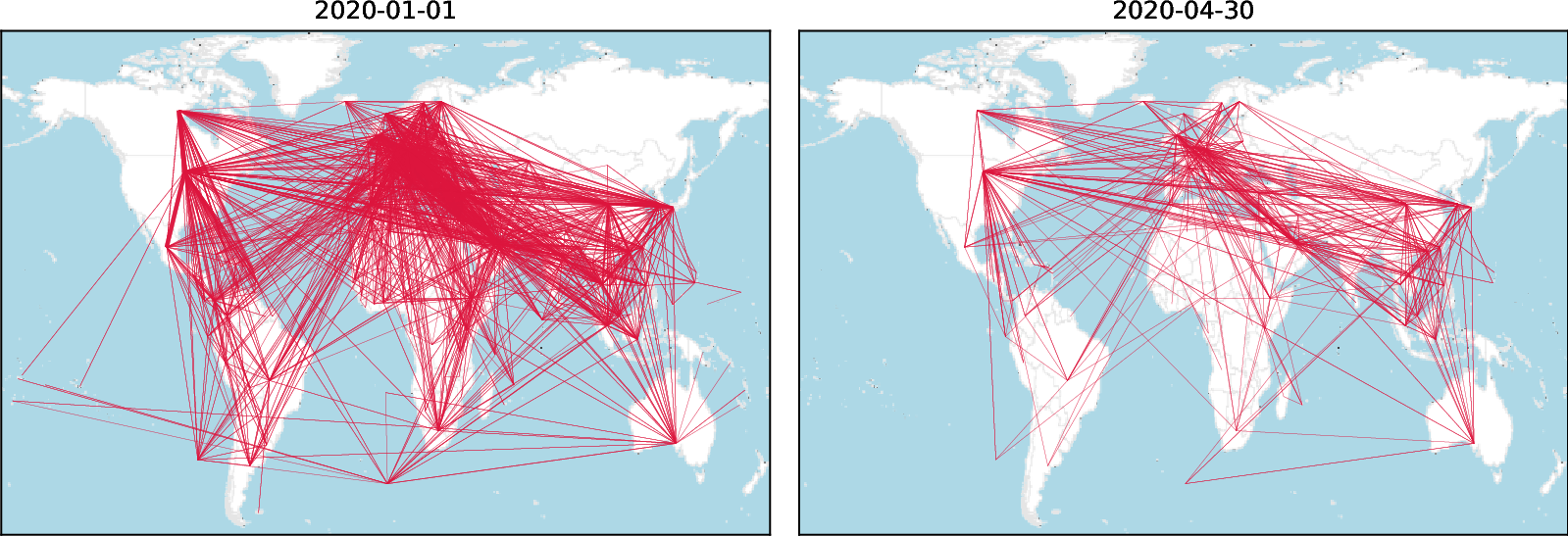}
\caption{Comparison of the air transportation country networks for January 1st, 2020 (left) and April 30th, 2020 (right). Nodes are countries and links represent international flights. A significant drop in the network connectivity can be observed throughout the COVID-19 pandemic.}
\label{fig:CountryNetworkSnapshots}
\end{figure}

\begin{figure}[b!]
\centering
\includegraphics[width=\textwidth]{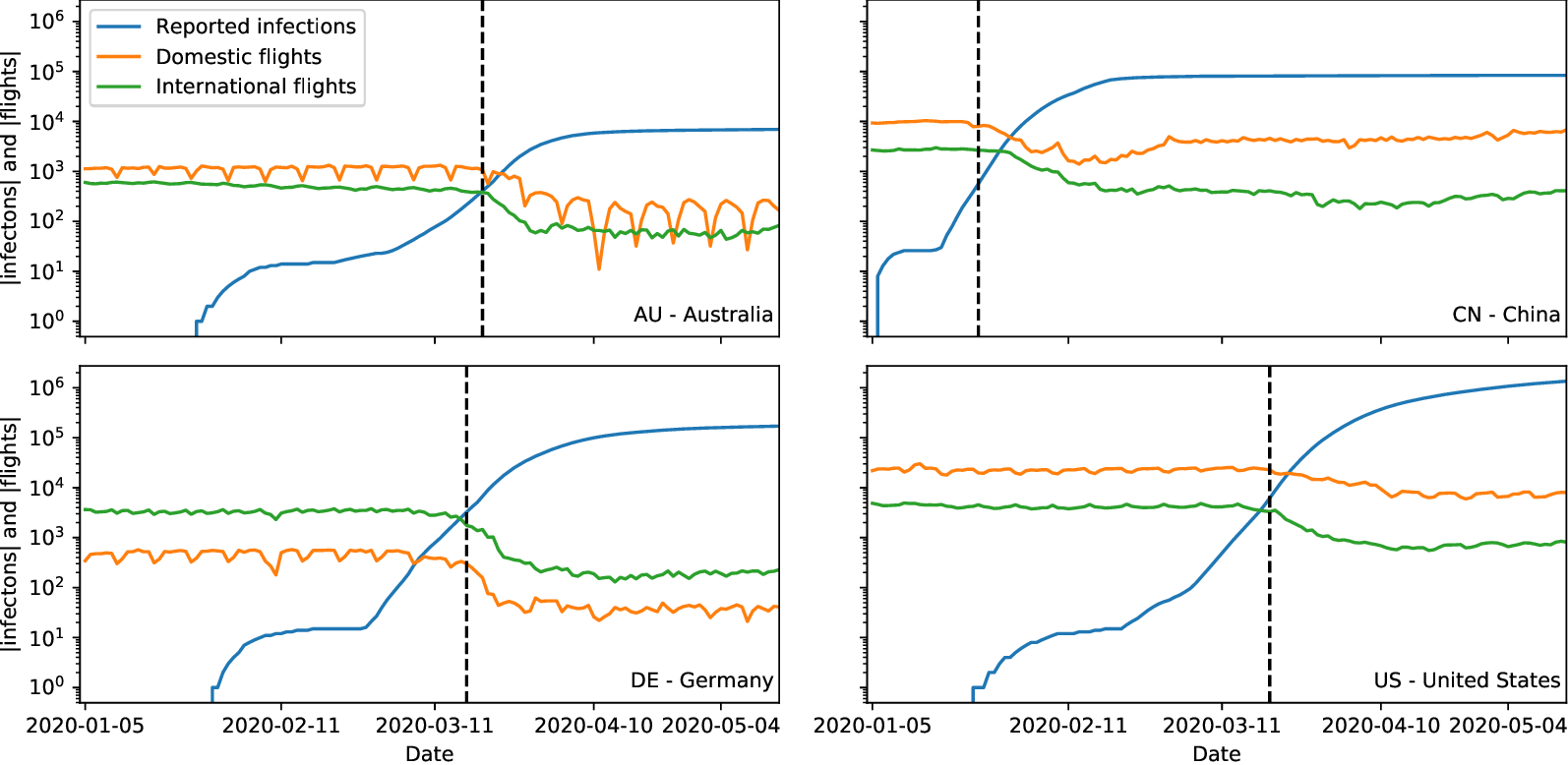}
\caption{A motivating example based on three indicators: the reported number of infections (curves in blue color), the number of domestic flights  (curves in orange color), and the number of international flights  (curves in green color), for four selected countries (Australia, China, Germany, and the United States.) The x-axis shows the timeline along the COVID-19 outbreak, the y-axis shows the number of infections, domestic flights, and international flights in a log scale, and the vertical dashed line represents the approximated time of drop in the number of international flights. 
}
\label{fig:MotivationExample}
\end{figure}

\begin{figure}[t!]
\centering
\includegraphics[width=\textwidth]{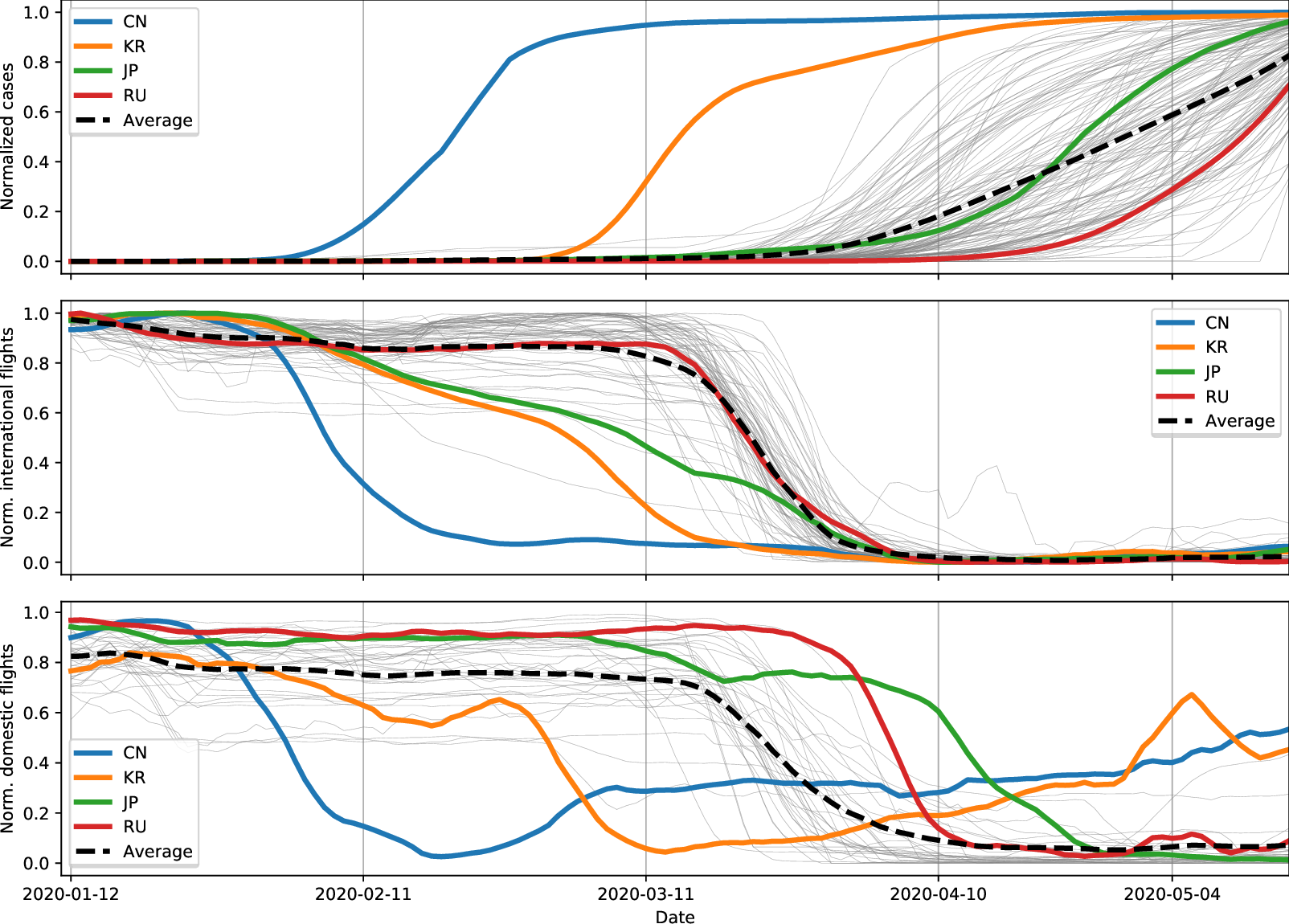}
\caption{The normalized number of infected cases for selected countries, the x-axis shows the timeline along the COVID-19 outbreak, while the y-axis shows the normalized number of infected cases, the normalized number of international flights, and the normalized number of domestic flights, ranging from 0.0 (minimum) to 1.0 (maximum). The light-grey curves denote the results of all remaining countries.
}
\label{fig:NormalizedAll}
\end{figure}

The data for this study was obtained from Flightradar24 and aggregated by airlines. In total, our dataset covers services of 150 airlines between 2,751 airports. All flights were grouped by days into 24-hour intervals; yielding one network per day induced by the flights taking place on that day. In total, we retrieved data for 152 days, from January 5th, 2020 to May 15th, 2020. The data about airports used in this study is from OurAirports (\url{https://ourairports.com/data/}), including airport codes, location and time zone. Figure~\ref{fig:CountryNetworkSnapshots} provides snapshot visualizations of the global air transportation country network  for two selected dates:  January 1st, 2020 (left) and April 30th, 2020 (right). It can be seen that the connectivity of the international country network drops significantly throughout the COVID-19 pandemic.

Furthermore, the impacts of COVID-19 on air transport are more profound on the international market than in the domestic market. This distinguishing effects can be seen more clearly in Figure~\ref{fig:MotivationExample}. In Figure~\ref{fig:MotivationExample}, we show an example of the three major criteria explored and analyzed in our study. For each country (Australia, China, Germany, and United States)\footnote{These four countries were selected to represent the effect pf the pandemic on distinct regions of the world, covering Asia, Europe, North America, and Oceania.}, the three indicators are plotted as a time series using a log scale. The first indicator is the cumulative number of reported infections in the country (blue color). All countries reveal a two-stage growth pattern, with an initial growth towards a few hundred cases, just to turn into a full-scale epidemic in each country within a few weeks; only the slope and the initial growth date are different among countries. The second indicator in our study is the number of international flights (green color). We can observe an expected seasonality of the number of flights; until a specific point where the number of international flights drops significantly. Similarly, the third indicator reports the number of domestic flights (orange color). The dashed, vertical line represents the approximate time of drop in the number of international flights for a country. 

For these motivating examples, it can be seen, that for all four countries, the second stage of the domestic epidemic was already reached when the airlines/governments decided to cancel these flights. This observation is striking, given that it is known for years, that air transportation plays a key role for turning a local epidemic into a global pandemic; because it allows long-distance travel in short amount of time; thereby reducing the effective distance between remote places. Given the limitation of the data, we cannot tell the load factors of aircraft which flew before the lock-down and its effect in the two indicators. It is reported that some airlines were flying (essentially) empty aircraft, in order to not lose their slots at the airports. Other airlines were operating off-schedule flights for conducting take-home missions of passengers stranded in remote places; even into times when travel bans were active. Nevertheless, the operations of these aircraft on its own, even if they were not fully loaded, puts the crews and therefore the destination countries at high risk. Therefore, these four examples motivate us to analyze the effect of synchronization between air transport operations and the increase in reported infections in this study.

The discussion of four example countries leads to the question whether these patterns are general and they holds for more countries. Accordingly, we investigate the normalized curves further in Figure~\ref{fig:NormalizedAll}. The data points for each country are normalized in a range from 0.0 (minimum) to 1.0 (maximum); a moving average filter with window size equal to seven was applied in order to remove outlier and reduce the effect of weekly seasonality; the window size of the moving average is set to seven. The following insights can be obtained from Figure~\ref{fig:NormalizedAll}. First of all, China and South Korea are outstanding from the other countries given the early rise in the number of reported cases Figure~\ref{fig:NormalizedAll}(top) as well as with the early reduction of the number of international flights and domestic flights Figure~\ref{fig:NormalizedAll}(middle, bottom). The vast majority of other countries has a significant delay in all three indicators.  

\begin{figure}[t!]
\centering
\includegraphics[width=\textwidth]{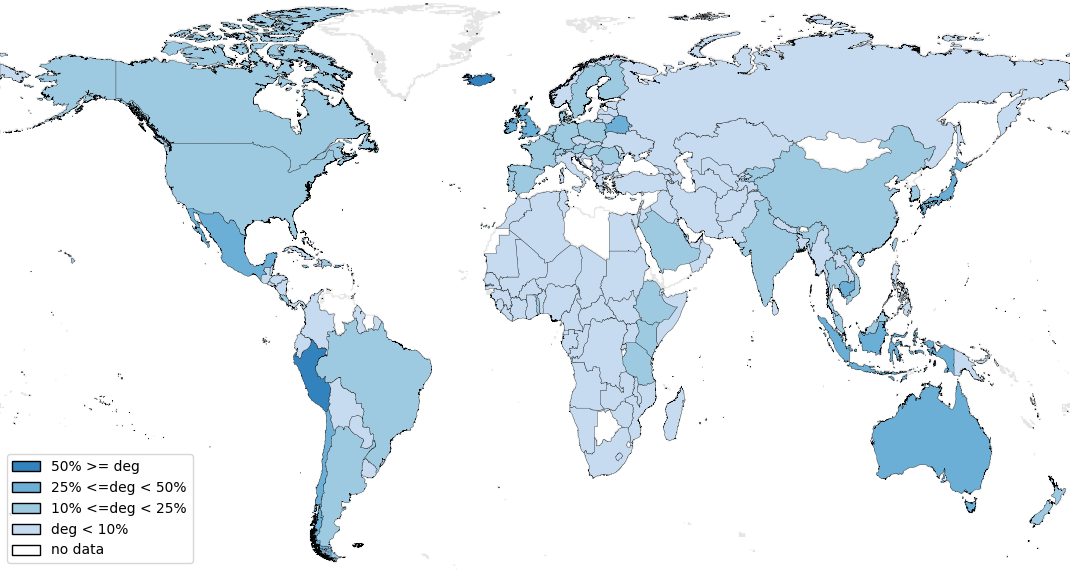}
\caption{Spatial distribution of countries with evolving connectivity patterns during the COVID-19 pandemic. Darker colors indicate less connectivity losses despite of the implementation of control measures all over the world. }
\label{fig:DegreeReductionMap}
\end{figure}

In the remaining part of this study, we will analyze this transition during the COVID-19 pandemic further; with an emphasis on the synchronization and type of events. First, we will take into account the degree, a network science node importance metric, which measures the number of direct neighbors. In our case, the degree of a country in the network corresponds to the number of destination countries reachable with a direct flight. Figure~\ref{fig:DegreeReductionMap} depicts the spatial distribution of countries and how they reacted to COVID-19. The color of a country corresponds to its degree on April 1st, 2020, divided by the maximum degree throughout the period of this study. It can be seen that the majority of countries have decreased their number of (country) destinations significantly during the pandemic; many of these countries below 25\%. A few countries stand out with remaining the degree of connectivity at a higher level, e.g., Mexico, Peru, Belarus, and Iceland.

\begin{figure}[t!]
\centering
\includegraphics[width=\textwidth]{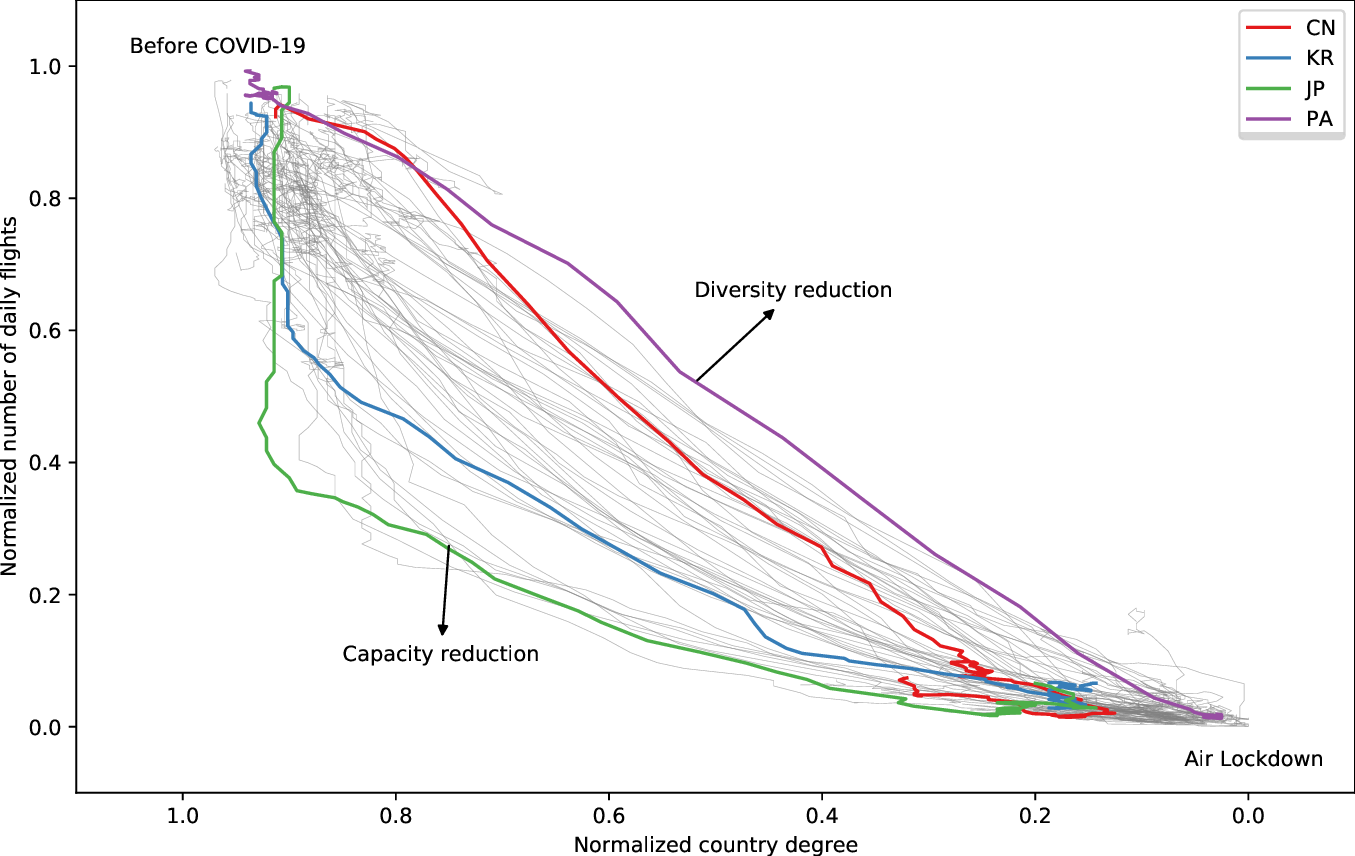}
\caption{The normalized degree (the number of flight connections with other countries) of a country versus the normalized number of international flights, where 0.0 denotes minimum and 1.0 denotes maximum after normalization. Four countries with representative patterns are highlighted: China, South Korea, Japan, and Panama. Two extremes of actions taken by different countries can be observed: Capacity reduction (represented by the green curve) and diversity reduction (represented by the purple curve). The light-grey curves denote the results of all remaining countries.
}
\label{fig:DegreeVSInternationalFlights}
\end{figure}

In the next experiment, we aim to understand the process of flight reductions with respect to the degree of a country. One can distinguish two possible extremes in such experiments. First, the so-called diversity reduction, in which the country attempts to remove the degree significantly; which means that the number of destination countries served by airlines is reduced. Second, the so-called capacity reduction, in which the countries keep most of their destinations alive, while reducing the frequency of connections significantly. Figure~\ref{fig:DegreeVSInternationalFlights} presents an analysis of the air transport lock-down strategy of individual countries. We plot the curves for the normalized degree over time (0.0=minimum degree of a country in the period, 1.0=maximum degree of a country in the period) against the normalized number of international flights. While there is no explicit temporal information in this chart, the countries, in general, move from the upper left (large degree, large number of flights) to the lower right (small degree, small number of flights). The chart reveals that both extremes, diversity reduction and capacity reduction, are present in the response to COVID-19. For instance, Japan, after a short period of fluctuation, decided to carefully reduce the number of daily flights, while maintaining a high degree of diversity; i.e., the fraction of flights was reduced by almost 60\%, but 80\% of the destination were still served; which means that capacity (or frequency) was the focus of the lock down. An example for a country towards the other extreme is Panama, which rather chose to cut all flights equally, sharply reducing its degree right from the start.

\begin{figure}[t!]
\centering
\includegraphics[width=\textwidth]{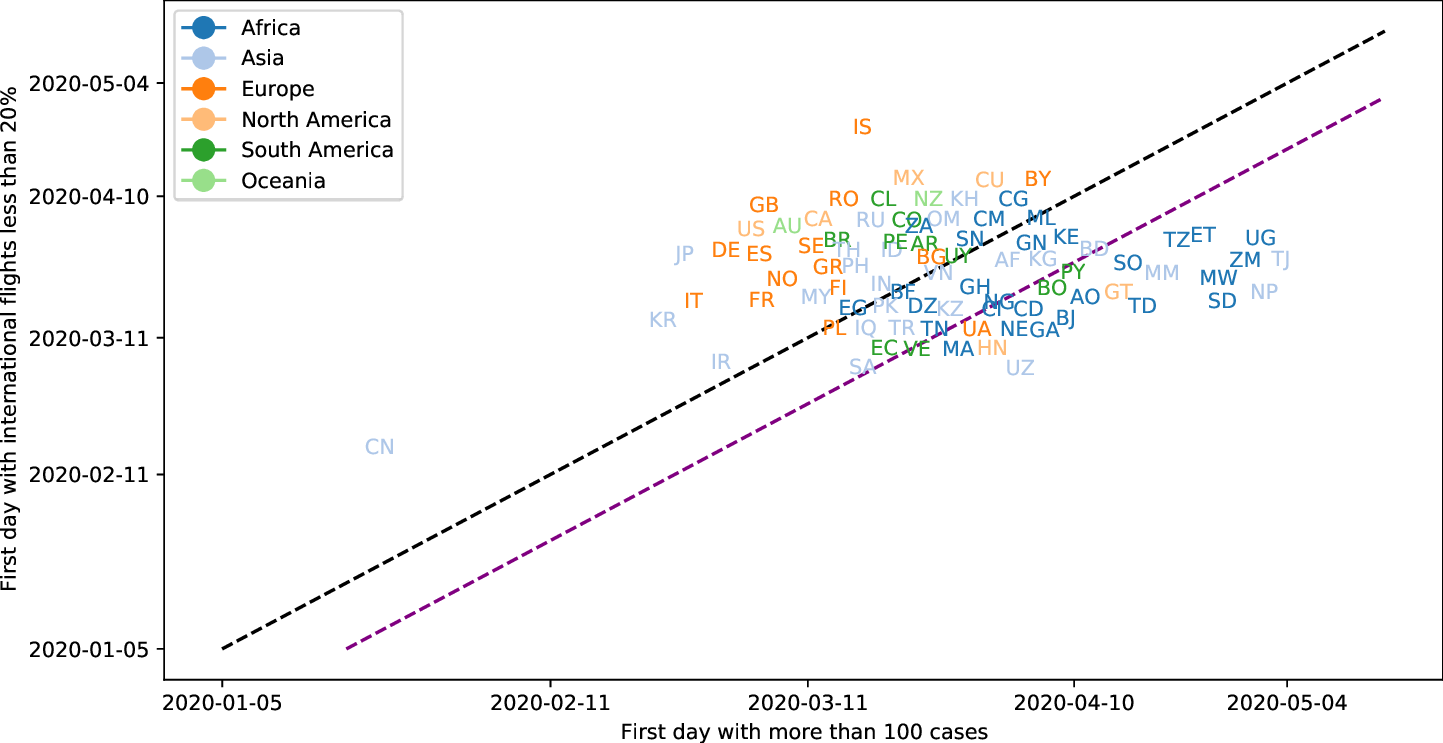}
\caption{Scatter plots for the day of the first 100 reported cases versus the first day with more than 80\% reductions in the number of international flights for different countries. Countries from different geographical locations are denoted with different colors. The black dashed diagonal line represents a perfect correlation between the occurrence of these two events; while the purple dashed diagonal line represents a 14-day pre-emptive flight reduction reference. Only data for countries with more than 10 international flights per day are shown; 90 countries matched the selection criterion.
}
\label{fig:InfectionsVSIntFlightReductions}
\end{figure}

Figure~\ref{fig:InfectionsVSIntFlightReductions} further pushes the envelope on answering the question whether the reaction of air transportation system was synchronized with the spread of the pandemic. For each country, we plot the day of the first 100 reported COVID-19 cases (x-axis) against the first day with more than 80\% reductions in the number of international flights for a country. The color encodes the continent of a country. The two diagonal lines represent the perfect correlation between both indicators (black) and a 14-day pre-emptive flight reduction baseline, in which countries have significantly reduced their air traffic 14 days before the outbreak of the first 100 cases (purple). Ideally, countries should be found towards the lower right of the purple diagonal line, if countries reacted early enough. Most of the countries, on the other hand, are located towards the upper-left, which means that the significant flight reduction happened much after reaching the first 100 domestic COVID-19 cases (we analyze the sensitivity of the number of cases below). Moreover, for the few countries located towards the lower-right, their timely flight reductions had all occurred after March 11, the date when WHO declared the disease to be a pandemic. In fact, the majority of these counties made the decision after April 10. This suggests some international learning has taken place at the country/airlines level. 

\begin{figure}[t!]
\centering
\includegraphics[width=\textwidth]{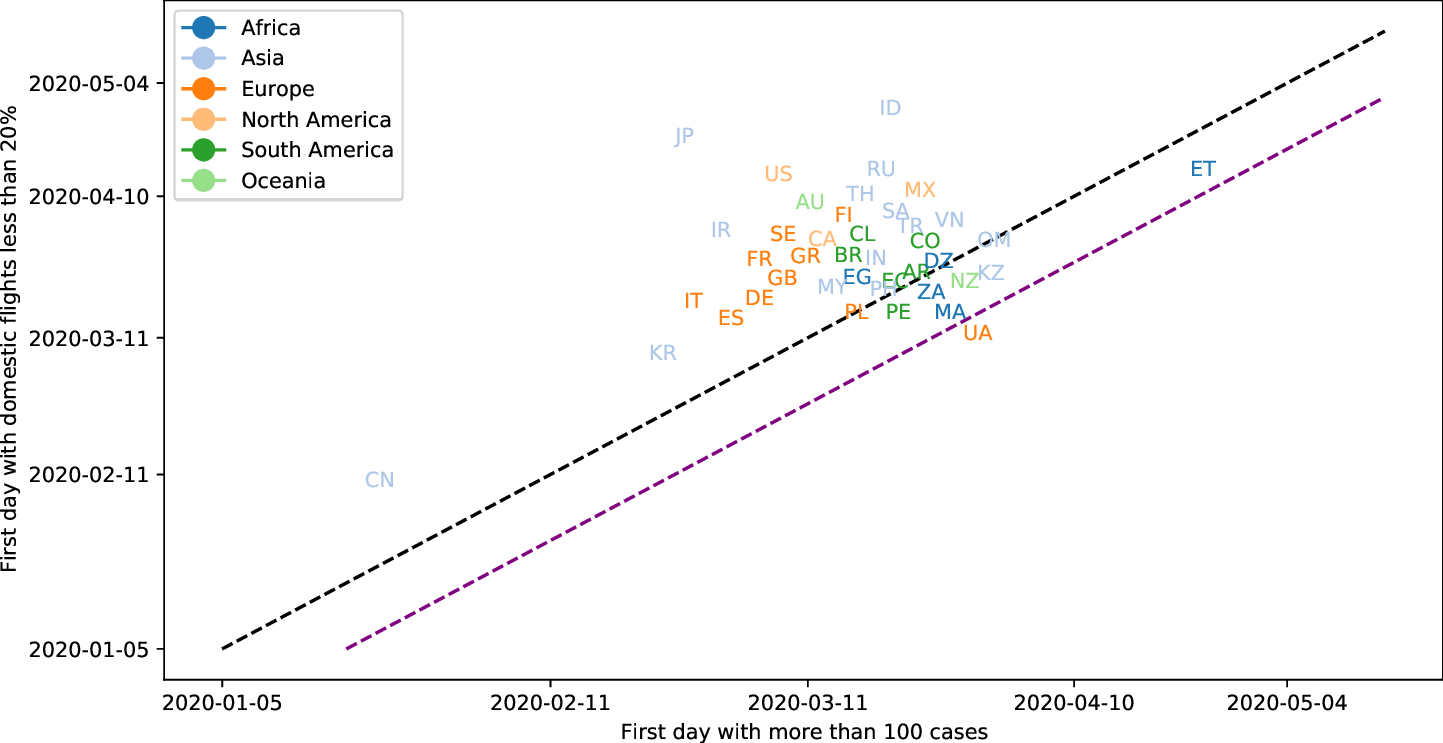}
\caption{Scatter plots for the day of the first 100 reported cases versus the first day with more than 80\% reductions in the number of domestic flights for different countries. Countries from different geographical locations are denoted with different colors. The black dashed diagonal line represents a perfect correlation between the occurrence of these two events; while the purple dashed diagonal line represents a 14-day pre-emptive flight reduction reference. Only data for countries with more than 10 domestic flights per day are shown; 42 countries matched the selection criterion.}
\label{fig:InfectionsVSDomFlightReductions}
\end{figure}

Given that distinguishing impacts of COVID-19 on the international market and the domestic market, we further examine the reactions of airlines on the domestic flights during the COVID-19 spread in Figure~\ref{fig:InfectionsVSDomFlightReductions}.
The day of the first 100 reported cases and the first day with only less than 20\% domestic flights connection remaining are compared and contrasted. 
While the general flight reduction decision is consistent with the case of the international flights in Figure~\ref{fig:InfectionsVSIntFlightReductions}, the actions of airlines concerning domestic flight operations are further delayed than those for international flight operations, this can be observed from the scatter plots that most data points appear above the black dashed diagonal line. 
Moreover, the number of countries taking actions on domestic flights is much less than those for international flights. 
This might be one indication that the deployment of cutting down international flights is more effective, and thus more widely implemented than reducing domestic flights, in addition to other anti-contagion measures, such as hygiene, social distancing, and home quarantine~\cite{Kraemer2020}. Figure~\ref{fig:InfectionsVSIntFlightReductionSensitivity} takes this analysis further, by presenting the number of reported, domestic cases by country, on the first day the international flights were reduced down to 20\%. It can be seen that those countries with a large total number of infections are also countries which had already a large number of cases reported; accordingly, one can conclude that from the analysis in this study, the measures might have been taken too late.

\section{Discussion and conclusions}
\label{sec:conclusions}

This study has investigated the degree of synchronization between air transport connectivity and the number of confirmed COVID-19 cases at a worldwide scale, with countries (governments) as individual decision makers. 
To the best of our knowledge, this study is the first to provide an empirical analysis on the impacts of COVID-19 on aviation at a country-level, using the most recent global air traffic data spanning over the last six months. We investigated the two-way relationships between air transportation and COVID: on one hand, air transport clearly affects the spread of COVID. While, on the other hand, we also investigated the impacts of COVID on aviation at worldwide level. The evolving dynamics of the synchronization relationships among three major criteria are reported: The number of reported infections, the number of international flights, and the number of domestic flights for several major countries. We summarize the major findings of our study below:

\begin{enumerate}

\item We observed that all countries reveal a two-stage pattern regarding the growth of the COVID-19 infected cases: While the initial growth towards a few hundred infections, within a few weeks the magnitude of infections quickly escalated to tens of thousands, turning the COVID-19 from a local epidemic to a pandemic in an extremely short of time. 

\item Strikingly, we found that almost all countries probably reacted simply too late in their decision to cut down flights, despite it is known for years that air transportation plays a key role in the spread of a pandemic. Therefore, the best strategic time window to implement control measures for the containment of COVID-19  is missed because of the delayed reactions from the governments (and airlines). Transportation lock-down measures work best before the virus is inside a country.

\begin{figure}[t!]
\centering
\includegraphics[width=\textwidth]{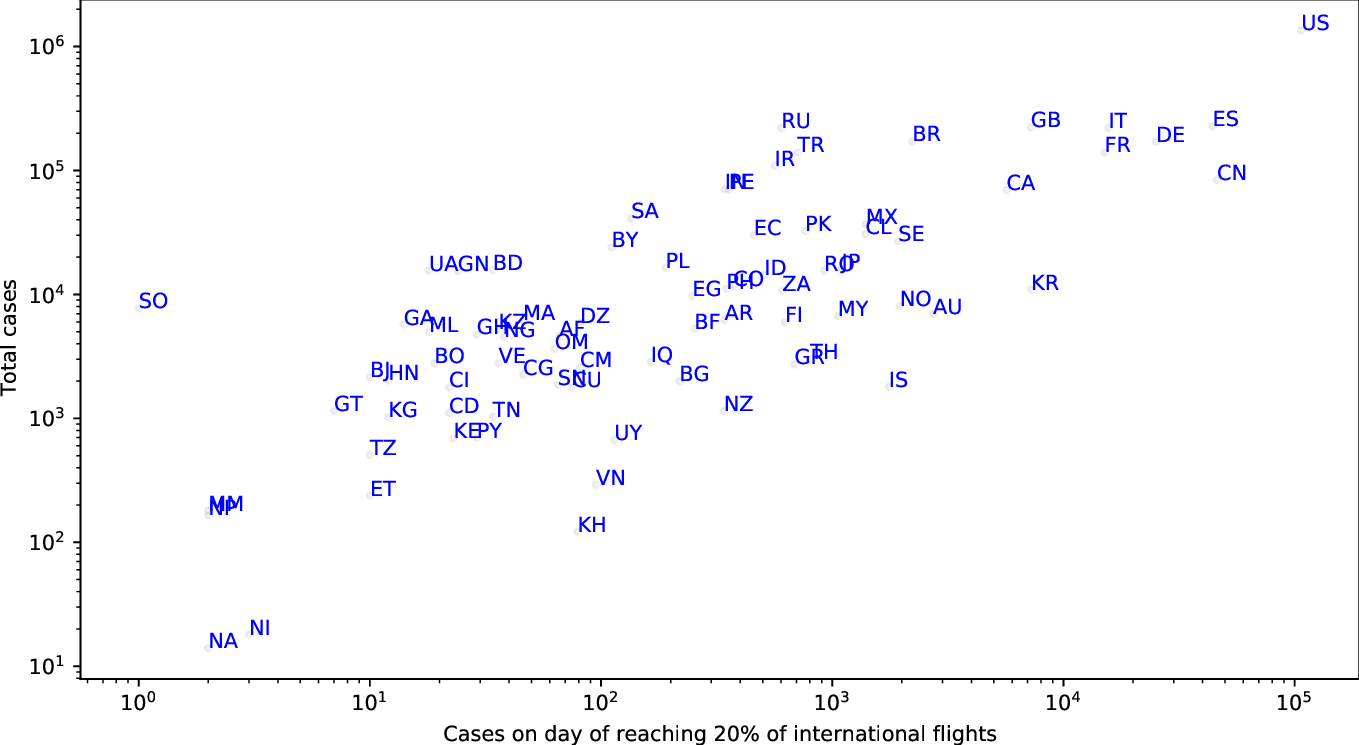}
\caption{The total number of cases in each country is plotted against the number of cases on the first day the number of international flights was reduced down to 20\% or below.}
\label{fig:InfectionsVSIntFlightReductionSensitivity}
\end{figure}

\item We applied a widely used complex network centrality metric, degree, which measures the number of direction connections a country has, to the international country network. Spatial distribution of countries with evolving connectivity patterns during the COVID-19 pandemic showed that the majority of countries have lately but significantly reduced the number of flight connections; many of these countries below 25\%. A few countries remain a high degree of connectivity such as Mexico, Peru, Belarus, and Iceland.

\item We observed two extremes of actions taken by different countries/airlines: Diversity reduction and capacity reduction. While the former strategy attempts to reduce degree significantly, i.e., the number of countries served by airlines; the latter strategy intends to reduce the frequency of connections, but keeping most of their connections alive. Results showed that both extremes of actions have been deployed by different countries in the response to COVID-19.

\item We also found that the reactions of airlines on the international flight operations and domestic flight operations are different, due to the distinguishing impacts of COVID-19 on these two markets. While the general flight reduction decision is consistent for airlines, the actions on domestic flights are even more delayed than those for international flights. This probably can be explained by the substantial effectiveness of cutting down international flights as one of the widely implemented anti-contagion policies. 

\end{enumerate}

Our study comes with a few set of limitations: First, we have analyzed the actual number of flights. This number does not contain any information on the load factor. It has been reported that some airlines were flying (essentially) empty aircraft, in order to not lose their slots at slot-constrained airports. Other airlines were performing off-schedule flights for conducting take-home missions of passengers stranded in remote places; even into times when travel bans were active. The take-home operations often had a load factor close to 100\%. Thus, the results of our study should be understood as a first step towards addressing the problem of synchronization. Nevertheless, the operations of these aircraft on its own, even if they were not fully loaded, puts the crews and therefore the destination countries at high risk. In addition, the cost incurred by operating empty flights (just for keeping slots) are tremendous. Finally, it has been reported that some carries have used their passenger aircraft for cargo 
transportation during some phases of the pandemic, in order to increase revenue. It is difficult to obtain such operational data from airlines directly; yet, it is a very interesting direction for future work to analyze these operations and their effect.

The results of the current study can facilitate the understanding of the roles of different countries during the COVID-19 pandemic, as well as the decision making of governments from regulators’ viewpoint and of international organizations (such as WHO, IATA, and ICAO) for better coordination and regulation. Future research could look at passenger behavior from the demand side: How does travel bans and mobility restrictions change the behavior of passengers, and thus the demand of air transportation systems, which would ultimately influence the airlines operations. Because of different business models, reactions at specific airline levels, in particular, traditional major airlines and newly emerged low-cost airlines, could be compared and contrasted, in order to recommend appropriate reaction strategies. Given limited airport slot resources and the motivation to minimize losses due to flight cancellations, it would be also interesting to look at the interplay between the operations of passenger flights and cargo flights for different airlines. 

\section*{Acknowledgement}
This study is supported by the National Natural Science Foundation of China (Grant No. 61861136005, No. 61650110516, and No. 71731001). 


\bibliographystyle{abbrv}
\bibliography{document}

\end{document}